\documentclass[preprint2]{emulateapj} 

\begin{document} 
 
\title{INFRARED OBSERVATIONS OF THE MILLISECOND PULSAR BINARY  
J1023+0038: evidence for short-term nature of its interacting phase 
in 2000--2001\footnote{{\it Herschel} is an ESA space observatory with  
science instruments provided by European-led Principal Investigator consortia  
and with important participation from NASA.}} 
 
\author{Xuebing Wang\altaffilmark{1,2}, 
Zhongxiang Wang\altaffilmark{1}, Nidia Morrell\altaffilmark{3}} 
 
\altaffiltext{1}{\footnotesize Key Laboratory for Research in Galaxies  
and Cosmology, Shanghai Astronomical Observatory, Chinese Academy of Sciences,  
80 Nandan Road, Shanghai 200030, China} 
 
\altaffiltext{2}{\footnotesize Graduate School of Chinese Academy of Sciences, 
No. 19A, Yuquan Road, Beijing 100049, 
China} 
 
\altaffiltext{3}{Las Campanas Observatory, Observatories of the Carnegie  
Institution of Washington, La Serena, Chile}

\begin{abstract} 
We report our multi-band infrared (IR) imaging of the transitional  
millisecond pulsar system J1023$+$0038, a rare pulsar binary known to have 
an accretion disk in 2000--2001. The observations  
were carried out with ground-based and space telescopes from near-IR  
to far-IR wavelengths.  We detected the source in near-IR $JH$ bands  
and \textit{Spitzer} 3.6 and 4.5 $\mu$m mid-IR channels. Combined with  
the previously-reported optical spectrum of the source, the IR emission  
is found to arise from the companion star, with no excess emission  
detected in the wavelength range. Because our near-IR fluxes are nearly  
equal to those obtained by the 2MASS all-sky survey in 2000 Feb., the 
result indicates that the binary did not contain the accretion disk at 
the time, whose existence would have  
raised the near-IR fluxes to 2-times larger values. Our observations  
have thus established the short-term nature of the interacting phase seen 
in 2000--2001: the accretion disk at most existed for 2.5 yrs. 
The binary was not detected by the WISE all-sky survey carried out in 2010  
at its 12 and 22 $\mu$m bands and our \textit{Herschel} far-IR imaging  
at 70 and 160 $\mu$m. Depending on the assumed properties of the dust,  
the resulting flux upper limits provide a constraint of  
$<3\times 10^{22}$--$3\times 10^{25}$~g on the mass of the dust grains  
that possibly  
exist as the remnant of the previously-seen accretion disk. 
 
\end{abstract} 
 
 
\section{INTRODUCTION} 
It was quite surprising when the binary millisecond pulsar (MSP) J1023+0038  
(hereafter J1023) was discovered in an untargeted radio pulsar survey  
\citep{arc+09}, because although the source was previously  
known to have radio emission \citep{bon+02}, it was suggested to be a  
cataclysmic variable (CV) and later has been established that it is likely  
a low-mass-X-ray binary (LMXB) with a 4.75 hour 
orbital period and a $\sim$0.2 $M_{\odot}$ companion  
\citep{ta05}.  
Detailed analyses of an optical spectrum of J1023, taken in 2001 February 
by the Sloan Digital Sky Survey (SDSS),  
have shown that the source at the time was in a bright state and had broad,  
double-peaked hydrogen and helium emission lines, a typical feature of  
an accretion disk  \citep{wan+09}.  
From 2002 May onward, however, the source has been seen dramatically different, 
having only a G-type absorption spectrum \citep{ta05} and 
thus indicating the disappearance of the disk approximately 1.5 yrs later.  
Combining with these observational results, the discovery of a radio pulsar  
has strongly suggested that this is the first such binary found at the end of  
its transition from a LMXB to a radio MSP. 
 
MSPs are believed to be "recycled" pulsars, formed from evolution of   
X-ray binaries \citep{bv91}. In such a binary 
with a low-mass companion,  
a neutron star was spun up by accretion of  
matter from the companion via an accretion disk. At the end of  
the interacting phase when the mass  
accretion rate in the disk was below a minimum value and thus 
the inner edge of the disk was truncated outside of the light cylinder of  
the neutron star by the neutron star's magnetic field, 
the neutron star appeared at radio frequencies as a MSP  
(\citealt{cam+98} and references therein).  
Subsequently, the disk would be disrupted due  
to the radiation pressure from the fast-spinning pulsar. While this 
model is generally 
accepted, there are few observational studies of the scenario due to the lack 
of known systems that are right at the end of the transition. 
 
From the estimated spin-down rate of the J1023 pulsar, its rotational energy  
loss rate (so called spin-down luminosity) has been found to be  
$L_{\rm sd}\simeq 4.3\times10^{34}$ ergs s$^{-1}$ \citep{del+12}.  
The major part of the energy should be 
carried out in a magnetized, high-velocity wind, which is likely 
interacting with the companion causing the orbital modulation \citep{ta05} and 
eclipses of the pulsar emission \citep{arc+09} seen  
at optical and radio frequencies, respectively. The wind also plausibly  
caused the disappearance of the disk that existed in 2000--2001  
(however see \citealt{tct10,tct12} for an alternative scenario). 
 
It is not clear whether the companion star in J1023 would replace 
its disk through repeatedly outflows.
If it does, as pointed out by 
\citet{ta05} that the Kelvin-Helmholtz relaxation time for the companion 
is much longer than 10 years since 2001 (see also \citealt{del+12}), 
J1023 would serve as a rare example for detailed studies of features of 
the transitional processes and pulsar-wind--disk interaction.  
 
In an effort to fully study this rare system and the detailed 
evolutionary process of such binary systems at or right after the transition 
phase, we have carried out infrared (IR) observations with 
ground-based and space telescopes. The accretion disk that has disappeared in 
the optical after 2002 May might have left an IR remnant. Recent  
\textit{Chandra} observations of the source have found orbital variability  
in its X-ray emission and likely revealed an intrabinary shock \citep{bog+11}. 
The shock, which would be produced by  
the interaction between the pulsar wind and outflow from the companion,  
might also be detectable at IR wavelengths. 
Our observations and data reduction are described in \S~\ref{sec:obs}.  
The results from the observations are given in \S~\ref{sec:res}, and 
the implications of our results for J1023 are discussed in \S~\ref{sec:dis}. 
 
\section{OBSERVATIONS} 
\label{sec:obs} 
 
\subsection{Near-IR Observation and Data Reduction} 
 
Our target was detected in near-IR $JHK_s$ bands by the 2MASS all-sky  
survey \citep{2mass} in 2000 February 6, but with relatively large  
uncertainties  
($J=16.30\pm 0.10$, $H=15.69\pm0.13$, and $K_s=15.9$ with no uncertainty given; 
\citealt{wan+09}).  In order to either confirm the 2MASS results or detect 
any variability,  we re-observed J1023 in $JH$  bands using  
the 2.5-meter Ir\'{e}n\'{e}e du Pont telescope at Las Campanas Observatory 
in Chile on 2012 May 10. The camera was the RetroCam, which uses a detector of 
a Rockwell Hawaii-1 $1024\times1024$ HgCdTe array. The field of view was 
$3.4\arcmin\times 3.4\arcmin$, with a pixel scale of 0.2\arcsec\ pixel$^{-1}$. 
The total on-source exposure time at each band was 405~s.  
During an exposure, the telescope was dithered in a 3$\times$3 grid with  
offsets of approximately 15\arcsec\ to obtain the measurement of  
the sky background.  The observing conditions were good, having  
0.6\arcsec\ seeing.  
 
We used the {\tt IRAF} data analysis package for data reduction. The images  
were dark-subtracted and flat-fielded. From each set of dithered images  
in one observation, a sky image was made by filtering out stars.  
The sky image was subtracted from the set of images, and  
then the sky-subtracted images were shifted and combined into one final  
image of the target field. 
 
The near-IR counterpart to J1023 was well detected in both J and H band images. 
We performed aperture photometry to measure brightnesses of the target and 
other in-field sources. 
Flux calibration was conducted by using the 2MASS sources detected in our 
images.  
 
\subsection{\textit{Spitzer} IRAC 3.6 and 4.5 $\mu$m Imaging} 
 
We observed the target on 2010 January 15 four times with  
the \textit{Spitzer} Space 
Telescope during its warm mission phase (Program ID 60197).  
The multiple exposures were 
requested in order to average 
out any possible orbital flux changes caused by the irradiated companion  
star by the pulsar. 
The imaging instrument used was the Infrared Array Camera (IRAC;  
\citealt{faz+04}). While the camera operated in four channels 
at 3.6, 4.5, 5.8, and 8.0 $\mu$m, only the first two channels were 
available during the warm mission and used for our observations. 
The detectors at the short and long wavelengths are InSb and Si:As devices, 
respectively, both with 256$\times$256 pixels and a plate scale  
of 1.2\arcsec\ pixel$^{-1}$. 
The field of view (FOV) is 5\arcmin$\times$5\arcmin. The frame time was 30 s 
with 23.6~s and 26.8~s effective integration time per frame, and 
the total integration times in each of our four observations 
were 1.97 min and 2.23 min at channel~1 and 2, respectively. 
 
The raw image data were processed through the IRAC data pipelines 
(version S18.13.0) at the \textit{Spitzer} Science Center (SSC). In the  
Basic Calibrated Data (BCD) pipeline, standard imaging data reductions,  
such as removal of 
the electronic bias, dark sky subtraction, flat-fielding, and linearization, 
are performed and individual flux-calibrated BCD frames are produced. 
The details of the data reduction in the pipelines can be found in 
the IRAC Instrument Handbook (version 2.0.2). 
Using a MOsaicker and Point source EXtractor (MOPEX) package, 
provided by the SSC, 
we conducted further data reduction and profile-fitting photometry. 
In the processes, we used the MOPEX pipelines {\tt Overlap} to perform 
background matching between the BCD frames, {\tt Mosaic} to carry out 
removal of cosmic rays and bad pixels, and {\tt APEX} to perform 
point source extraction. 
Also using {\tt Mosaic}, final post-BCD (PBCD) mosaics were produced 
by combining the BCD frames. The target was well detected in our  
IRAC images. 
 
\subsection{WISE Imaging} 
 
Launched on 2009 December 14, the Wide-field Infrared Survey Explorer (WISE)  
mapped the entire sky at 3.4, 4.6, 12, and 22 $\mu$m (called W1,  
W2, W3, and W4 bands, respectively) in 2010 with FWHMs of  
6.1\arcsec, 6.4\arcsec, 6.5\arcsec, and 12.0\arcsec\  in the four bands,  
respectively \citep{wri+10}.  
The WISE all-sky images and source catalogue were released in 2012 March. 
We checked the WISE data and found that J1023 was detected in  
W1 and W2 bands but not in the other two bands. 
We downloaded the images of the target field from 
the Infrared Processing and Analysis Center (IPAC).  
The dates of the observations and the depth of coverage of the target field  
were between 2010 May 21--26 and 11--12 pixels (corresponding to 97--106~s  
on-source integration time), respectively.  
 
\subsection{\textit{Herschel} PACS 70 and 160 $\mu$m Imaging} 
 
We also observed J1023 on 2011 November 26 with the \textit{Herschel}  
Space Observatory (Obs. ID 1342233048, 1342233049). 
The imaging instrument used was the Photodetector Array Camera and  
Spectrometer (PACS; \citealt{pog+10}). 
PACS performs imaging at a blue and a red channel simultaneously, 
of which the central wavelengths are either 70 or 100 and 
160 $\mu$m, respectively. The detectors are a 64$\times$32 and 
a 32$\times$16 bolometer array for the blue and red channels, respectively, 
both with a field of view of $\sim 1.75\arcmin\times 3.5\arcmin$. 
We used the mini-scan map mode of PACS for imaging at  
the channels of 70 $\mu$ (bandwidth 25 $\mu$m) and  
160 $\mu$m (bandwidth 85 $\mu$m). The total telescope time for our 
observation was 1 hour with 18~min on-source time. 
 
The image data were downloaded from the \textit{Herschel} Science Archive, which 
were processed from the \textit{Herschel} Data 
Processing system at the \textit{Herschel} Science Centre (HSC). The target was 
not detected in the images. We used the background counts at the source region 
and estimated the flux upper limits. The values we derived were highly  
consistent with that given by the \textit{Herschel} observation planning tool 
{\tt HSpot}. 
 
\section{Results} 
\label{sec:res} 
 
The obtained flux measurements for J1023 from 2MASS $JHK_s$,  
our near-IR $JH$, and \textit{Spitzer}/IRAC 3.6 and 4.5 $\mu$m detections 
are listed in Table~\ref{tab:sum}. We assigned a 0.3~mag uncertainty to 
the 2MASS $K_s$ band measurement. 
The $JH$ magnitudes resulting from our observation are consistent  
with the 2MASS values 
within the uncertainties. For the \textit{Spitzer} observations at each  
IRAC channel, our four flux measurements showed $\leq$1.2\% and $\leq$3.4\%
differences at 3.6 and 4.5 $\mu$m, respectively, and the averaged values are
given in the table. Considering 2.7\% and 3.5\% uncertainties on the
measurements from photometry at the two channels, no significant flux
variations were detected. We also used the orbital period and time of 
ascending node provided from radio timing of the pulsar \citep{arc+09}, and
checked the phase duration of our \textit{Spitzer} observations. 
The phase was in a range of 0.58--0.67, corresponding to 0.33--0.42 
used in \citet{ta05}, during which the optical light curve 
(emission from the companion) of J1023 has a flat top at $V$ band and a
$V-I$ color variation of $\lesssim$0.02 mag (see Figure 4 in \citealt{ta05}). 
Since the effective temperature of the companion was probably around 
5700~K \citep{ta05} at the time of our \textit{Spitzer} observations, 
the expected orbital flux variations should be smaller than
0.02~mag at the mid-IR wavelengths, indicating that our result
of no significant flux variations is consistent. 
\begin{center} 
\includegraphics[scale=0.8]{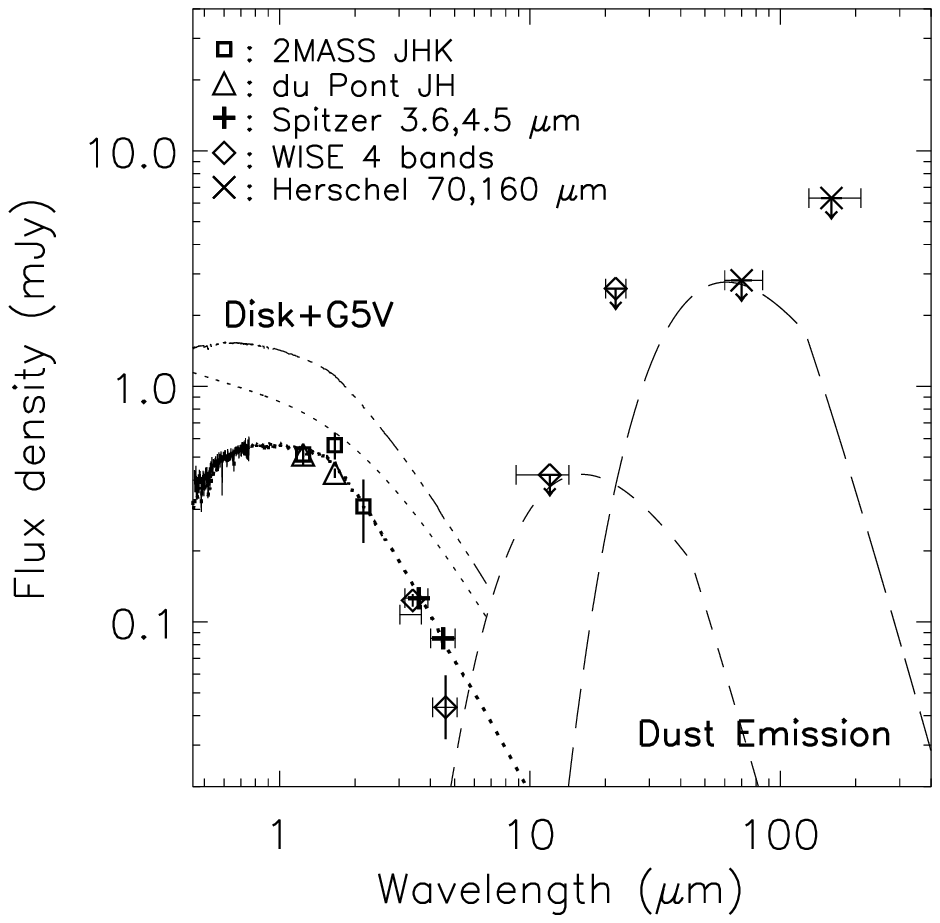} 
\figcaption{IR flux measurements and upper limits of J1023, with
the horizontal bars showing the mid- and far-IR band-passes.
The averaged optical spectrum from \citet{ta05} is 
plotted as a solid curve. A G5V model spectrum (dotted curve), which
is scaled by a radius of 0.4 $R_{\sun}$ at a distance of 1.37 kpc,
well fit the optical spectrum and near-IR and mid-IR flux measurements, 
indicating  non-detection 
of excess IR emission at the wavelengths. A model spectrum of  
the accretion disk in 2001 and the combined spectrum of the disk and 
companion are over-plotted (dash dotted and long-dash dotted curves, 
respectively). Model spectra from two sizes 
of dust grains are shown (dashed and long-dashed curves are for 
7 and 20 $\mu$m  dust grains, respectively), providing upper limits 
on the dust mass of  
$3\times 10^{22}$ and $3\times 10^{25}$~g for the first and latter size dust. 
\label{fig:flux} } 
\end{center} 

The WISE 3.4 and 4.6 $\mu$m flux measurements  
from the WISE catalogue are also provided in Table~\ref{tab:sum}. 
For the non-detections of J1023 in WISE W3 and W4 bands and 
our \textit{Herschel} images, flux upper limits were obtained 
and their 3$\sigma$ values are given.
The WISE measurements of J1023 in W1 and W2 bands,  
particularly the latter, 
have magnitudes larger than those of the IRAC (by 0.127 and 0.68 mag), 
although the differences are within 3$\sigma$ 
uncertainties of the WISE measurements. It has been noted that at the faint
end starting from $>$15 mag at W1 and W2 bands, there is a faintward
bias (in average of 10--20\%) of WISE relative to IRAC measurements 
from the comparisons of photometry of sources in the ecliptic poles
\citep{jar+11}.\footnote{see also http://wise2.ipac.caltech.edu/docs/release/allsky\\
/expsup/sec6\_3c.html}
This bias is suspected to be caused by stronger source confusion to WISE due to
its relatively lower resolution. Also the comparisons show $\sim$0.1 and 
$\sim$0.2 mag ranges of scattering of WISE W1 and W2 measurements, 
respectively, when sources are faint with $>$15 mag. 
Therefore we concluded that the apparent differences of the WISE measurements 
of J1023 relative to that of our IRAC are instrumental, due to 
the bias and larger scattering of
WISE W1 and W2 measurements of relatively faint sources. 
 
\subsection{Near-IR and Mid-IR detections} 
 
Optical emission from J1023 after 2002 May is known to arise 
from the companion star, which has a G5V type optical spectrum and probably 
has a size approximately equal to that of the Roche lobe (0.4 $R_{\sun}$;  
\citealt{ta05, wan+09}). We display the optical spectrum, the same 
one given in \citet{wan+09} which was averaged from 23 spectra 
obtained in 2003-2004 by \citet{ta05}, in Figure~1. We found that the spectrum
can be well fit with a G5V model spectrum \citep{kur93} with no
reddening needed ($\chi^2=857$ for 1498 degrees of freedom).
We note that in \textit{Chandra} X-ray spectroscopic studies of the source,
no hydorgen column density was needed either \citep{bog+11}. At a distance
of $d=1.37$ kpc \citep{del+12}, the companion's radius $R$ was found to be
$R\approx 0.41\ R_{\sun}$.
The flux densities at near-IR 1.24 and 1.66 $\mu$m and 
mid-IR 3.6 and 4.5 $\mu$m wavelengths were thus derived from the model 
spectrum and they are 0.498, 0.437, 0.119, and 0.076 mJy, respectively. 
The model flux densities are consistent with the observed values within
1$\sigma$ uncertainties at near-IR wavelengths 
and slightly lower than that at mid-IR wavelengths (Table~2). 
Adding 2\% systematic
uncertainties at IRAC channels 1 and 2 \citep{rea+05}, the values are
consistent within approximately 2$\sigma$ uncertainties.
We thus conclude that near-IR and mid-IR emission we detected  
likely arose from the companion star and no significant excess IR emission 
was detected from the source.

We estimated flux upper limits on excess mid-IR emission from J1023.
Considering that near-IR and mid-IR emission detected in our ground-based
and \textit{Spitzer} observations
arose from the companion, we fit them and the optical spectrum together 
with the same model spectrum used above.
In order to reduce the dominance of the optical spectrum,  
we assigned a weight of 1/150 to them. 
In addition, 2\% systematic uncertainty from flux calibration at near-IR
wavelengths and the systematic uncertainties of IRAC channels 1 and 2 
were included.
We found that $R\approx 0.42\ R_{\sun}$ at $d=1.37$~kpc provided
the best-fit, and the minimum $\chi^2=14.9$ for 13 degrees of freedom.
Assuming a detection of mid-IR emission at either 3.6 or 4.5 $\mu$m
with the same uncertainty as that in our \textit{Spitzer} observations 
and fixing $R=0.42\ R_{\sun}$, we increased 
the flux value at each wavelength till that $\chi^2$ was increased by 8.9 
(3$\sigma$ confidence; \citealt{lmb76}). The
3$\sigma$ upper limits were thus found to be 0.011 and 0.006 mJy at
channels 1 and 2, respectively.
  
\subsection{Implication of the near-IR measurements} 
 
It was noted by \citet{wan+09} that the 2MASS measurements of J1023 made 
on 2000 Feb. 6 were consistent with being at the tail of  
the G5V spectrum of the companion. Combining with the mid-IR measurements, 
our more accurate near-IR $JH$ measurements, which 
are approximately equal to the 2MASS values, confirm the 2MASS detection 
and the companion-star origin of the near-IR emission detected both in 2000 and 
at the current time. This result raises a very interesting point that 
in 2000 Feb. the accretion disk did not exist. In Figure~\ref{fig:flux} we 
over-plotted the disk model spectrum shown in \citet{wan+09} 
and the combined spectrum of the disk plus 0.4~$R_{\sun}$ G5V star. 
As can be seen, 
if the accretion disk had existed in 2000 Feb., the near-IR fluxes would 
have been approximately 2 times larger, $J=15.2$, $H=15.0$, and $K_s=14.8$, 
based on the combined model spectrum.  
 
\citet{bon+02} presented an optical spectrum of J1023, which was taken on 2000 
May 6 and contains the same emission features as the SDSS spectrum.  
The existence of the accretion disk thus likely started after 2000 Feb. 6 but 
before May 6. A summary of the properties of J1023 known from different
observations around the time is given
in Table~1 (also see \citealt{arc+09}). 
Previously \citet{wan+09} found a required time  of 0.5 yrs  
for building 
the accretion disk in 2001 by considering the estimated disk mass of  
10$^{23}$~g and mass transfer rate of 10$^{16}$ g~s$^{-1}$.  
Our constraint on the starting time of the disk existence 
is therefore roughly consistent with their estimation. The life 
time of the interacting phase should be at most 2.5 yrs from 2000 Feb. 
to 2002 May, suggesting that for J1023 the interacting phase is of short-term 
in nature and that the system might repeat the short-term process in 
the future when the outflow from the companion could overcome the pulsar wind. 
Close monitoring of the binary system is warranted. 
 
\section{Discussion} 
\label{sec:dis} 
 
We have conducted multi-band imaging observations of J1023 from near-
to far-IR wavelengths using ground-based and space telescopes. In the near-IR 
$JH$ bands and \textit{Spitzer} mid-IR 3.6 and 4.5 $\mu$m channels,  
we have detected  
the source but the emission has been found to arise from the companion star  
of the binary and no excess IR emission was detected.  
The flux upper limits on the existence of any hot (600--800~K) dust  
in the system were derived and they were in a range of 0.006--0.011 mJy
at mid-IR 3.6--4.5 $\mu$m wavelengths.
At the longer wavelengths,  
we did not detect emission from the source and the 3$\sigma$ 
upper limits were approximately from 0.5 mJy at 12 $\mu$m to 6~mJy at 
160 $\mu$m. 
 
Constraints on the existence of relatively cold dust material in the binary system can be  
estimated from our results. A remnant of the accretion disk that was  
disrupted by either the pulsar wind \citep{wan+09} or $\gamma$-ray emission  
from the pulsar \citep{tct10,tct12} might stay in the system and be illuminated 
by the pulsar wind. We used a simple dust model given by \citet{ff96}, in which 
a single temperature is assumed for all dust grains, 
to estimate infrared emission from possible cold dust material in J1023.  
The spin-down luminosity of the pulsar in J1023 was recently estimated 
by \citet{del+12}. Considering that a fraction of this energy output $f=0.01$ 
heats the dust of two sizes, 7 and 20 $\mu$m, we 
found that our IR upper limits constrain the masses of the dust to be lower 
than 
3$\times 10^{22} (f_{0.01}/a_7 T_{\rm h}^6)$~g and 
3$\times 10^{25} (f_{0.01}/a_{20} T_{\rm l})^6$~g, 
where $a_7$ and $a_{20}$ are the dust grain sizes of 7 and 20~$\mu$m,
respectively. The corresponding 
dust temperatures are $T_{\rm h}\sim 300$~K and $T_{\rm l}\sim 80$~K. 
The model spectra are 
displayed in Figure~\ref{fig:flux}. 
We note that \citet{wan+09} have estimated 10$^{23}$~g mass for the accretion  
disk seen in 2001. Presuming that 1\% of the remnant's mass is the dust,  
the upper  
limits on the total mass of the remnant are $\sim$$10^{24}$~g for 
the dust of 300~K or $\sim$$10^{26}$~g for the dust of 80~K. The masses are  
substantially higher than the mass of the accretion disk, suggesting that  
the observations were not deep enough (particularly at 8 $\mu$m with no 
\textit{Spitzer} observations) for  
the detection of the possibly existing dust material in the system. 
 
X-ray emission from J1023 has been detected with a dominant non-thermal  
component \citep{hom+06,arc+10,bog+11}. Combined with the fact that  
orbital variability was seen in the X-ray emission, the origin of  
the emission is likely  
an intrabinary shock, produced by the outflows from the two stars of  
the system \citep{bog+11,tct12}. This shock could radiate emission 
in a long wavelength range from X-ray to IR wavelengths  
(see, e.g., \citealt{sla+08}). 
However, assuming a single power-law emission and extending  
the \textit{Chandra} 
spectrum ($\alpha=-0.2$; \citealt{bog+11}) to the IR wavelengths, a flux 
at 10 $\mu$m is predicted to be 10$^{-4}$ mJy. Therefore much deeper  
mid- or far-IR observations than ours are needed in order to determine  
possible emission properties 
of the putative intrabinary shock at the long wavelengths.   
 
\acknowledgements 
 
This publication makes use of data products from the Two Micron All Sky Survey, 
which is a joint project of the University of Massachusetts and  
the Infrared Processing and Analysis Center/California Institute of  
Technology, funded by the National Aeronautics and Space Administration  
and the National Science Foundation.   
The publication also makes use of data products from the Wide-field Infrared  
Survey Explorer, which is a joint project of the University of California,  
Los Angeles, and the Jet Propulsion Laboratory/California Institute of  
Technology, funded by NASA. 
This work is based in part on  
observations made with the \textit{Spitzer} Space  
Telescope, which is operated by the Jet Propulsion Laboratory, California  
Institute of Technology under a contract with NASA. 
 
We thank the anonymous referee for helpful comments.
This research was supported by 
National Basic Research Program of China 
(973 Project 2009CB824800), and National Natural Science 
Foundation of China (11073042). 
ZW is a Research Fellow of the  
One-Hundred-Talents project of Chinese Academy of Sciences. 
 
{\it Facility:} \facility{Ir\'{e}n\'{e}e du Pont (RetroCam), \textit{Spitzer} (IRAC), \textit{Herschel} (PACS)} 
 

\begin{thebibliography}{18}
\expandafter\ifx\csname natexlab\endcsname\relax\def\natexlab#1{#1}\fi

\bibitem[{{Archibald} {et~al.}(2010){Archibald}, {Kaspi}, {Bogdanov},
  {Hessels}, {Stairs}, {Ransom}, \& {McLaughlin}}]{arc+10}
{Archibald}, A.~M., {Kaspi}, V.~M., {Bogdanov}, S., {Hessels}, J.~W.~T.,
  {Stairs}, I.~H., {Ransom}, S.~M., \& {McLaughlin}, M.~A. 2010, \apj, 722, 88

\bibitem[{{Archibald} {et~al.}(2009)}]{arc+09}
{Archibald}, A.~M., {et~al.} 2009, Science, 324, 1411

\bibitem[{{Bhattacharya} \& {van den Heuvel}(1991)}]{bv91}
{Bhattacharya}, D., \& {van den Heuvel}, E.~P.~J. 1991, \physrep, 203, 1

\bibitem[{{Bogdanov} {et~al.}(2011){Bogdanov}, {Archibald}, {Hessels}, {Kaspi},
  {Lorimer}, {McLaughlin}, {Ransom}, \& {Stairs}}]{bog+11}
{Bogdanov}, S., {Archibald}, A.~M., {Hessels}, J.~W.~T., {Kaspi}, V.~M.,
  {Lorimer}, D., {McLaughlin}, M.~A., {Ransom}, S.~M., \& {Stairs}, I.~H. 2011,
  \apj, 742, 97

\bibitem[{{Bond} {et~al.}(2002){Bond}, {White}, {Becker}, \&
  {O'Brien}}]{bon+02}
{Bond}, H.~E., {White}, R.~L., {Becker}, R.~H., \& {O'Brien}, M.~S. 2002,
  \pasp, 114, 1359

\bibitem[{{Campana} {et~al.}(1998){Campana}, {Colpi}, {Mereghetti}, {Stella},
  \& {Tavani}}]{cam+98}
{Campana}, S., {Colpi}, M., {Mereghetti}, S., {Stella}, L., \& {Tavani}, M.
  1998, \aapr, 8, 279

\bibitem[{{Deller} {et~al.}(2012)}]{del+12}
{Deller}, A.~T., {et~al.} 2012, \apjl, 756, L25

\bibitem[{{Fazio} {et~al.}(2004)}]{faz+04}
{Fazio}, G.~G., {et~al.} 2004, \apjs, 154, 10

\bibitem[{{Foster} \& {Fischer}(1996)}]{ff96}
{Foster}, R.~S., \& {Fischer}, J. 1996, \apj, 460, 902

\bibitem[{{Homer} {et~al.}(2006){Homer}, {Szkody}, {Chen}, {Henden}, {Schmidt},
  {Anderson}, {Silvestri}, \& {Brinkmann}}]{hom+06}
{Homer}, L., {Szkody}, P., {Chen}, B., {Henden}, A., {Schmidt}, G., {Anderson},
  S.~F., {Silvestri}, N.~M., \& {Brinkmann}, J. 2006, \aj, 131, 562

\bibitem[{{Jarrett} {et~al.}(2011)}]{jar+11}
Jarrett, T. H., et al. 2011, ApJ, 735, 112

\bibitem[{{Kurucz} (1993)}]{kur93}
Kurucz, R. L. 1993, Kurucz CD-ROM, Cambridge, MA: Smithsonian Astrophysical Observatory

\bibitem[{{Lampton} {Margon} \& {Bowyer}(1976)}]{lmb76}
Lampton, M., Margon, B., Bowyer, S. 1976, ApJ, 208, 177

\bibitem[{{Poglitsch} {et~al.}(2010)}]{pog+10}
{Poglitsch}, A., {et~al.} 2010, \aap, 518, L2

\bibitem[{{Reach} {et~al.}(2005)}]{rea+05}
Reach, W. T., et al. 2005, PASP, 117, 978

\bibitem[{{Skrutskie} {et~al.}(2006)}]{2mass}
{Skrutskie}, M.~F., {et~al.} 2006, \aj, 131, 1163

\bibitem[{{Slane} {et~al.}(2008){Slane}, {Helfand}, {Reynolds}, {Gaensler},
  {Lemiere}, \& {Wang}}]{sla+08}
{Slane}, P., {Helfand}, D.~J., {Reynolds}, S.~P., {Gaensler}, B.~M., {Lemiere},
  A., \& {Wang}, Z. 2008, \apjl, 676, L33

\bibitem[{{Szkody} {et~al.}(2003)}]{szk+03}
Szkody, P., et al. 2003, AJ, 126, 1499

\bibitem[{{Takata} {et~al.}(2010){Takata}, {Cheng}, \& {Taam}}]{tct10}
{Takata}, J., {Cheng}, K.~S., \& {Taam}, R.~E. 2010, \apjl, 723, L68

\bibitem[{{Takata} {et~al.}(2012){Takata}, {Cheng}, \& {Taam}}]{tct12}
---. 2012, \apj, 745, 100

\bibitem[{{Thorstensen} \& {Armstrong}(2005)}]{ta05}
{Thorstensen}, J.~R., \& {Armstrong}, E. 2005, \aj, 130, 759

\bibitem[{{Wang} {et~al.}(2009){Wang}, {Archibald}, {Thorstensen}, {Kaspi},
  {Lorimer}, {Stairs}, \& {Ransom}}]{wan+09}
{Wang}, Z., {Archibald}, A.~M., {Thorstensen}, J.~R., {Kaspi}, V.~M.,
  {Lorimer}, D.~R., {Stairs}, I., \& {Ransom}, S.~M. 2009, \apj, 703, 2017

\bibitem[{{Wright} {et~al.}(2010)}]{wri+10}
{Wright}, E.~L., {et~al.} 2010, \aj, 140, 1868

\end{thebibliography}

\clearpage
\begin{deluxetable}{lllc}
\tablecolumns{4}
\tablecaption{Observational properties of J1023 around 
2000--2001\label{tab:prop}}
\tablewidth{0pt}
\tablehead{
\colhead{Date} & \colhead{Measurement} & \colhead{Inference} &  
\colhead{Reference}  }
\startdata
1998 Aug 10 & $F_{1.4 GHz}=$6.56 (mJy) & Pulsar emission (?) & 1\\
1999 Mar 22 & SDSS photometry & G dwarf color & 1,2\\
2000 Feb 6  & 2MASS near-IR photometry & G dwarf fluxes & 3,4\\
2000 May 6 & Optical spectroscopy & Disk present & 1\\
2001 Feb 1 & SDSS spectroscopy & Disk present & 3\\ 
2001 Dec 10 & Optical spectroscopy & Disk present & 2\\
2002 May 11 & Spectropolarimetry & G dwarf spectrum & 5\\
2003 Jan 31 & Optical spectroscopy & G dwarf spectrum & 6\\
2004 Jan 18--20 & Optical spectroscopy & G dwarf spectrum & 6\\
2004 Mar 8-9 & Optical spectroscopy & G dwarf spectrum & 6\\
2004 Nov 18 & Optical spectroscopy & G dwarf spectrum & 6\\
2004 Feb 16 & Spectropolarimetry & G dwarf spectrum & 5\\
2005 May 23 & Optical spectroscopy & G dwarf spectrum & 5\\
2007 Jun 28 & $F_{0.35 GHz}\simeq 75$ (mJy) & Pulsar discovery & 7\\
2008 Dec 23 & Optical spectroscopy & G dwarf spectrum & 7\\
2008 Dec 25 & Optical spectroscopy & G dwarf spectrum & 7\\
\enddata 
\tablerefs{
(1) \citet{bon+02};
(2) \citet{szk+03};
(3) \citet{wan+09};
(4) this work;
(5) \citet{hom+06};
(6) \citet{ta05}
(7) \citet{arc+09}.}
\end{deluxetable} 
 
\clearpage 
 
\begin{deluxetable}{llccc} 
\tablecolumns{4} 
\tablecaption{Flux measurements and upper limits of J1023\label{tab:sum}} 
\tablewidth{0pt} 
\tablehead{ 
\colhead{Telescope/Instrument} & \colhead{Observation date} &  
\colhead{Band} &  \colhead{Magnitude}  & \colhead{Flux (mJy)}} 
\startdata 
2MASS & 2000 Feb 6 & $J$ & 16.30$\pm$0.10 & 0.481$\pm$0.045 \\ 
      &            & $H$ & 15.69$\pm$0.13 & 0.542$\pm$0.065 \\ 
      &            & $K_s$ & 15.9$\pm$0.3\tablenotemark{a} & 0.29$\pm$0.08 \\  
du Pont/RetroCam & 2012 May 10 &  $J$ & 16.260$\pm$0.024 & 0.499$\pm$0.011 \\ 
                 &             & $H$  & 15.964$\pm$0.025 & 0.421$\pm$0.019 \\ 
Spitzer/IRAC & 2010 Jan 15 & 3.6 & 15.862$\pm$0.034 & 0.127$\pm$0.004 \\ 
             &             & 4.5 & 15.813$\pm$0.038 & 0.085$\pm$0.003 \\ 
WISE & 2010 May 21--26 & 3.4 & 15.989$\pm$0.066 & 0.123$\pm$0.008 \\ 
      &                    & 4.6 & 16.49$\pm$0.34 & 0.043$\pm$0.014 \\ 
     &                     & 12 &  $<$12.1 & $<$0.4 \\ 
     &                     & 22 & $<$8.8 & $<$2.6 \\ 
Herschel/PACS & 2011 Nov 26 & 70 &   & $<$2.8 \\ 
              &             & 160 &   & $<$6.3 \\ 
\enddata 
\tablenotetext{a}{0.3 mag uncertainty is assigned by us to $K_s$} 
\tablecomments{For non-detections, 3$\sigma$ flux upper limits are given.} 
\end{deluxetable}

\end{document}